# A Novel Decoupling Method for Investigating Distinct Domain Evolution in Ferroelectric Film

Keying Sun*, Jiajun Qiu*, Hao Li, Zongwei Shang, Lang Zeng#, Ming Li, Lining Zhang#, Runsheng Wang#

***Abstract*** — Conventional polarization characterization techniques provide only the averaged response of ferroelectric films, limiting the investigation of domain-dependent reliability mechanisms in $HfO_2$-based ferroelectrics. In this work, a new domain decoupling method is proposed to separately analyze ferroelectric domains with distinct switching behaviors. A three-domain model consisting of upward non-switchable domains (**P**↑), downward non-switchable domains (**P**↓), and switchable domains (**P**$_d$) is first introduced to describe heterogeneous domain populations during electrical cycling. By combining complementary switching-current measurements, the responses of different domain populations can be selectively extracted and reconstructed. The proposed method enables quantitative tracking of individual domain populations during cycling. As a demonstration, the method is further applied to analyze the time-dependent evolution of domain populations after electrical cycling. This approach provides a new route for investigating imprint, fatigue, and other reliability-related phenomena in ferroelectric devices.

## I. Introduction

$HfO_2$-based ferroelectrics have attracted significant attention in recent years due to their great CMOS compatibility and tremendous potential for applications in non-volatile memory [1-10]. However, the reliability issues remain significant challenges for practical applications. For the investigation of critical cycling-related reliability issues, including the wake-up and fatigue effect, standard *P-V* hysteresis loops and PUND measurements have been widely adopted to characterize the evolution of remnant polarization ($P_r$) [11-20]. However, these conventional methods only capture the global average polarization response of the entire ferroelectric film, and fail to provide additional details on domains with distinct polarization states. This limitation has severely hindered further research on the intrinsic reliability mechanisms of $HfO_2$-based ferroelectric.

The key challenge is therefore not merely measuring the total polarization, but resolving how different domain populations contribute to it during electrical cycling. In this work, we address this issue by tracking switching-current signatures associated with distinct domain states. A conceptual framework is established to correlate different domain states with their switching-current signatures, and a set of tailored electrical sequences is developed to probe their responses selectively. This strategy decomposes the collective ferroelectric response into domain-specific contributions, providing direct access to the evolution of individual domain populations beyond conventional polarization measurements.

This work was supported by NSFC (Grant No. T2293703, T2293700, 62125401, 62171009, 62425406), Beijing Outstanding Young Scientist Program (JWZQ20240101004) and the 111 Project (B18001).

Keying Sun and Lang Zeng are with Fert Beijing Institute, MIIT Key Laboratory of Spintronics, School of Integrated Circuit Science and Engineering, Beihang University, 100191, Beijing, China. Lang Zeng is also with the National Key Laboratory of Spintronics, Hangzhou International Innovation Institute, Beihang University, 311115, Hangzhou, China. (Email: zenglang@buaa.edu.cn)

Jiajun Qiu and Lining Zhang are with School of Electronic and Computer Engineering, Peking University, Shenzhen, 518055, China. (Email: lnzhang@ieee.org)

Jiajun Qiu, Hao Li, Zongwei Shang, Ming Li and Runsheng Wang are with the School of Integrated Circuits, Peking University, 100871, Beijing, China. (Email: r.wang@pku.edu.cn)

Keying Sun and Jiajun Qiu contributed equally to this work.

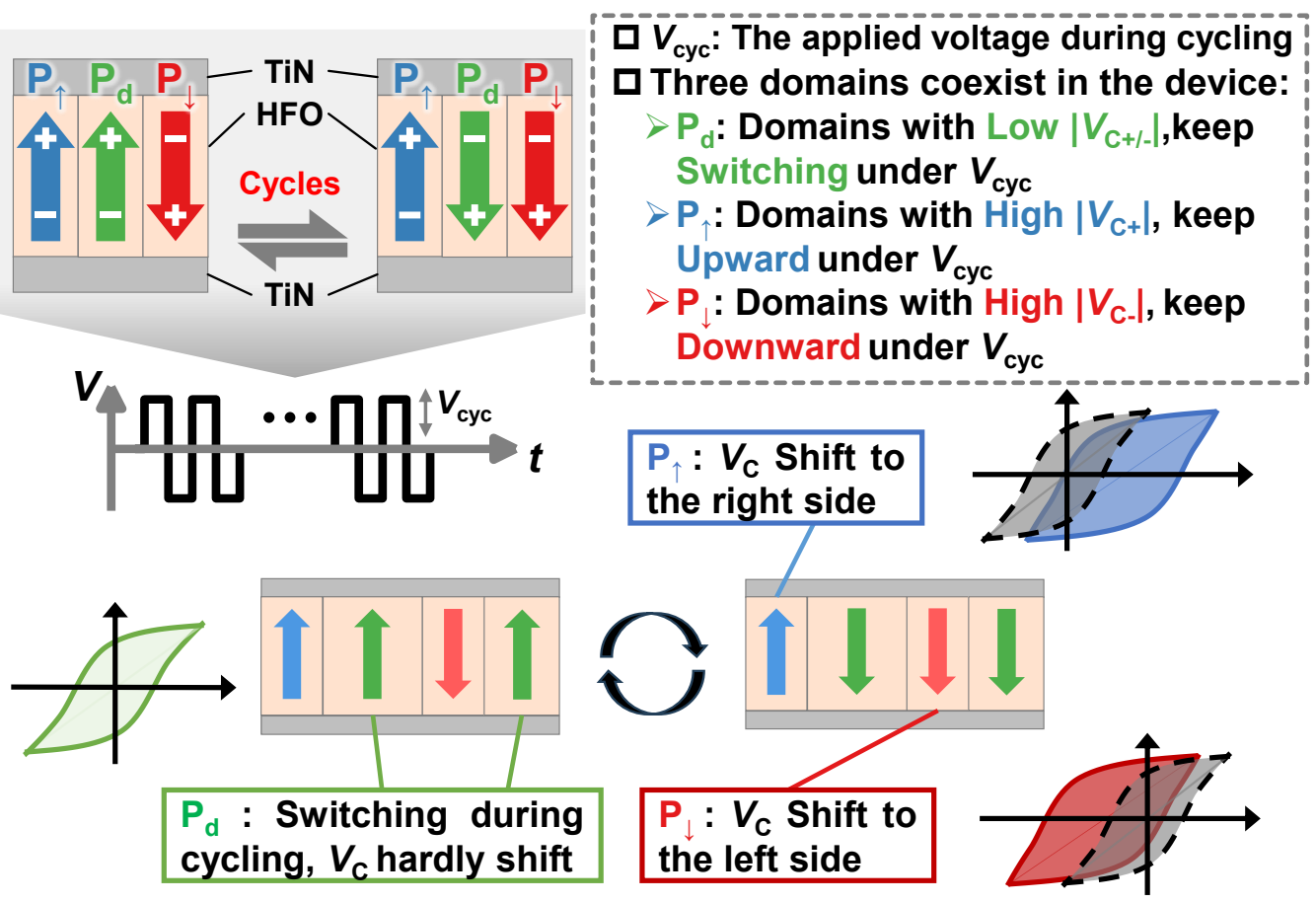

Fig. 1 Schematic of three ferroelectric domain under cyclic voltage in ferroelectric film: switchable domains (**P**$_d$, green) that flip with minimal $V_{C+}$ shift, upward non-switchable domains (**P**$_↑$, blue) with $V_{C+}$ shifted right, and downward non-switchable domains (**P**$_↓$, red) with $V_{C-}$ shifted left, illustrating their distinct switching behavior.

## II. Concept and Principle of Domain Decoupling

**Fig. 1** illustrates the conceptual framework of the proposed domain model. Under repeated bipolar voltage cycling, ferroelectric domains do not respond uniformly due to the broad distribution of coercive voltages ($V_C$) induced by defect accumulation or local internal fields [27,28]. As a result, ferroelectric domains can be divided into three components depending on their switching state during cycling, as shown in **Fig. 1**. We first define **P**$_↑$/**P**$_↓$ as those domains remaining unswitchable and stable upward/downward polarization state, as their positive/negative coercive voltage ($V_{C+}$ / $V_{C-}$) exceeds the applied pulse amplitude. **P**$_d$ is defined as those dynamic domains of continually switching in cycling, whose $V_{C+}$ and $V_{C-}$ are lower than the amplitude of cycles. **P**$_d$ is further defined as **P**$_{d,↑}$ and **P**$_{d,↓}$ by its retained upward or downward polarization state, which is determined by the polarity of the final pulse in the cycle sequence. Consequently, the measured switching current after cycling is a superposition of contributions from these three domain populations. By selectively controlling the polarization states of **P**$_d$, **P**$_↑$, and **P**$_↓$ through specifically designed voltage sequences, their individual switching-current components can be isolated and quantitatively extracted. This three-domain model forms the physical basis of the proposed domain decoupling methodology.

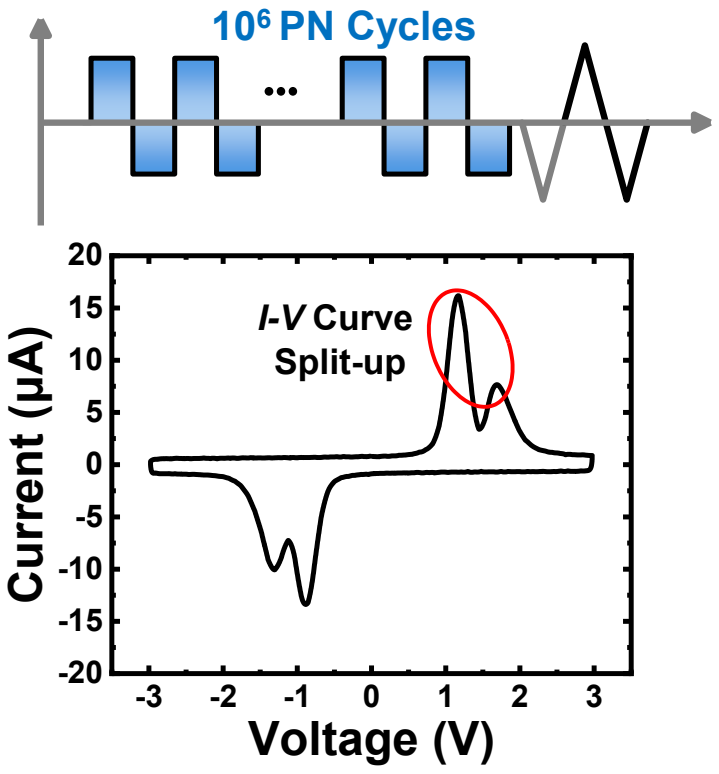

Fig. 2 Waveform and corresponding *I-V* curve. The switching peak splits into multiple components, indicating the coexistence of ferroelectric domain populations with distinct switching behavior.

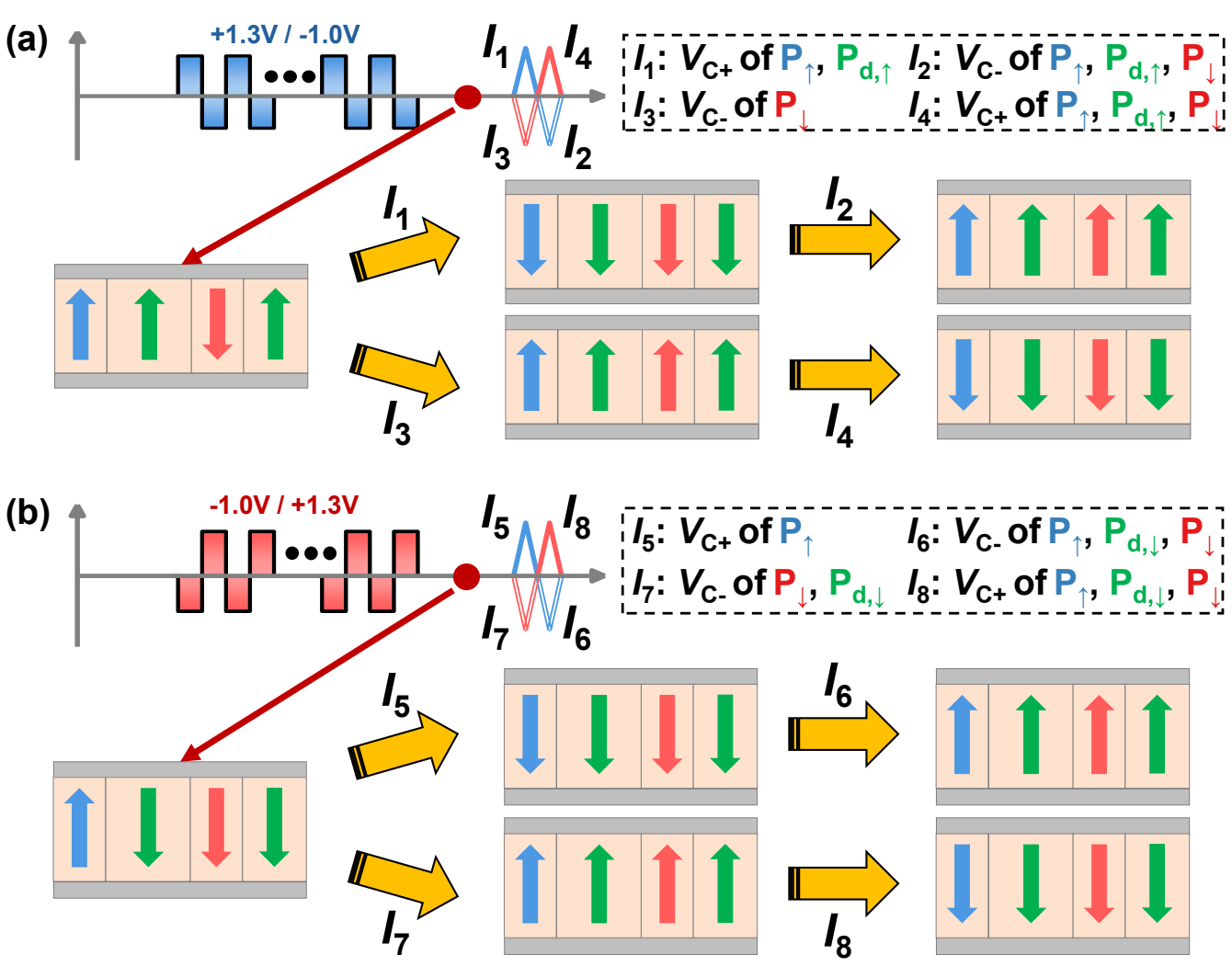

Fig. 3 Domain-state evolution and switching-current composition under two asymmetric cycling schemes: (a) positive-negative cycling (+1.3 V / –1.0 V) and (b) negative-positive cycling (–1.0 V / +1.3 V). The resulting switching-current peaks (***I*₁-*I*₈**) contain different combinations of **P**↑, **P**↓, and **P**d, providing the basis for domain decomposition.

## III. Experimental Demonstration of Domain Decoupling

The process flow and fundamental properties of the devices measured in this work are detailed in [24]. To investigate the switching behavior of different domain populations, the devices were first cycled using a reduced voltage amplitude and subsequently characterized by a larger-amplitude triangular voltage sweep to record the switching current. The resulting *I-V* curve is shown in **Fig. 2**. A clear splitting of the switching peaks is observed after cycling, indicating the coexistence of multiple domain populations with different coercive-voltage distributions. This observation is consistent with the three-domain model proposed in **Fig. 1**, where the coercive voltage of $\mathbf{P}_{\uparrow}$ shifts toward the positive voltage side, while that of $\mathbf{P}_{\downarrow}$ shifts toward the negative voltage side, while $\mathbf{P}_{d}$ remains nearly unchanged. Consequently, the overall switching current can be resolved into three distinct peak components. Such peak splitting has been widely reported in ferroelectric films and is commonly associated with domain pinning induced by defect accumulation and internal bias fields during electrical cycling [28]. However, the individual contributions of these domain populations remain

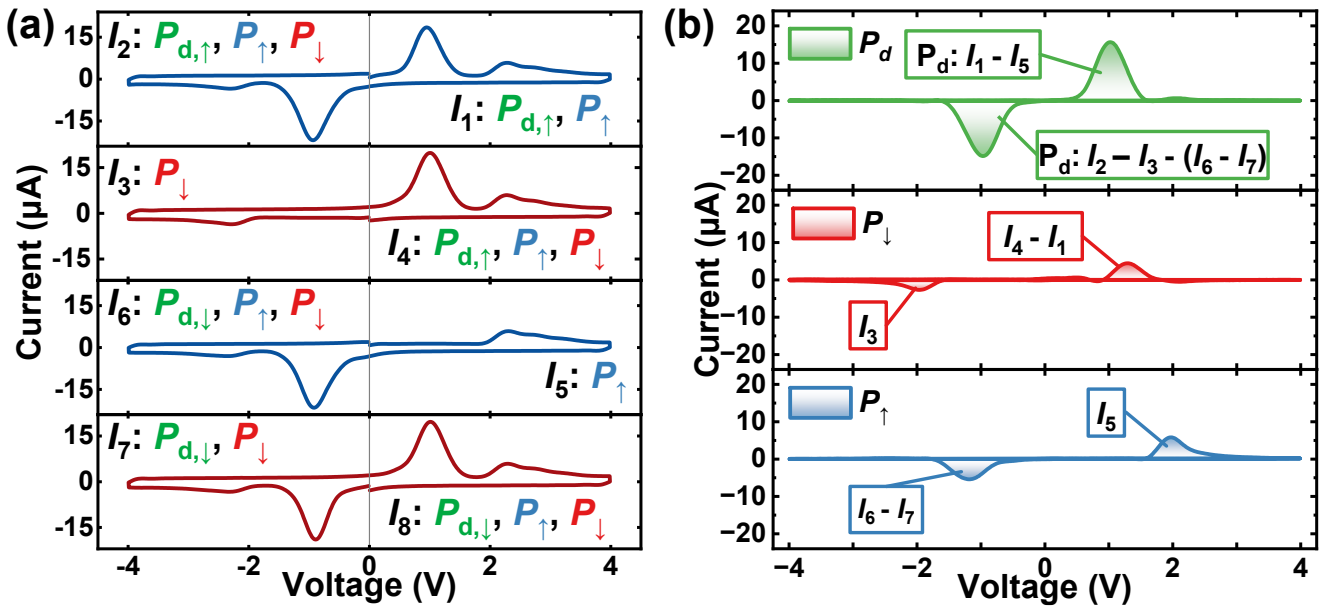

Fig. 4 (a) Detailed switching-current components measured under different cycling conditions, showing mixed contributions from **P**↑, **P**↓, and **P**d in the raw signals (***I*₁-*I*₈**). (b) Intermediate subtraction results of the measured current components, where specific combinations of signals (e.g., ***I*₄–*I*₁**, ***I*₆–*I*₇**) partially isolate distinct domain contributions, providing the basis for subsequent full decoupling of three ferroelectric domain populations.

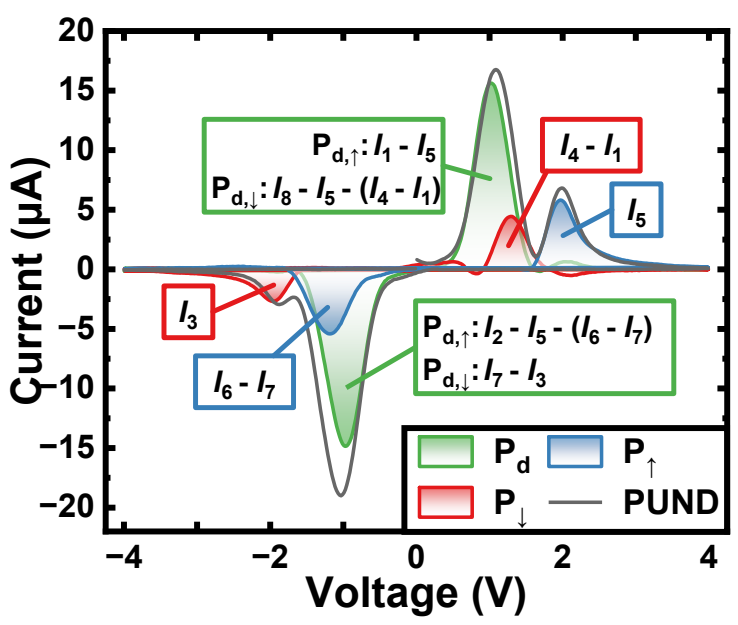

Fig. 5 Reconstructed switching-current spectra of the three ferroelectric domain populations obtained using the domain-decoupling procedure. The individual contributions from **P**↑, **P**↓, and **P**d are fully separated, revealing their distinct distributions and confirming the effectiveness of the proposed decomposition approach.

mixed in the measured *I-V* curve and cannot be directly quantified.

As illustrated in **Fig. 3**, *I-V* curves with distinct contributions from these components can be obtained by applying two sweep waveforms with opposite sequences after cycling waveforms with varying orders. For the waveform terminated with a negative pulse, $\boldsymbol{I_1}$ contains the contributions from $\mathbf{P}_{\uparrow}$ and $\mathbf{P}_{d,\uparrow}$ in the positive voltage segment; $\boldsymbol{I_2}$ contains the contributions from $\mathbf{P}_{\uparrow}$, $\mathbf{P}_{\downarrow}$, and $\mathbf{P}_{d,\uparrow}$ in the negative voltage segment; $\boldsymbol{I_3}$ contains the contribution from $\mathbf{P}_{\downarrow}$ in the negative voltage segment; and $\boldsymbol{I_4}$ contains the contributions from $\mathbf{P}_{\uparrow}$, $\mathbf{P}_{\downarrow}$, and $\mathbf{P}_{d,\uparrow}$ in the positive voltage segment. For the waveform with identical amplitude but terminated with a positive pulse, $\boldsymbol{I_5}$ contains the contribution from $\mathbf{P}_{\uparrow}$ in the positive voltage segment, $\boldsymbol{I_6}$ contains the contributions from $\mathbf{P}_{\uparrow}$, $\mathbf{P}_{\downarrow}$, and $\mathbf{P}_{d,\downarrow}$ in the negative voltage segment, $\boldsymbol{I_7}$ contains the contributions from $\mathbf{P}_{\downarrow}$ and $\mathbf{P}_{d,\downarrow}$ in the negative voltage segment and $\boldsymbol{I_8}$ contains the contributions from $\mathbf{P}_{\uparrow}$, $\mathbf{P}_{\downarrow}$ , and $\mathbf{P}_{d,\downarrow}$ in the positive voltage segment. Based on these domain-selective measurements, the individual switching-current spectra can be reconstructed through a series of subtraction operations, as shown in **Fig. 4** and **Fig. 5**. The contributions from $\mathbf{P}_{\uparrow}$, $\mathbf{P}_{\downarrow}$, and $\mathbf{P}_{d}$ are independently extracted, and the corresponding polarization can be calculated by integrating the decoupled current signals. Using the reconstructed domain contributions, the evolution of $\mathbf{P}_{\uparrow}$, $\mathbf{P}_{\downarrow}$ and $\mathbf{P}_{d}$ during cycling can be further analyzed.

As an application example, the proposed domain decoupling method was employed to investigate the evolution of domain populations after electrical cycling. **Fig. 6** presents the measured *I-V* curves at different delay times and the corresponding decoupled domain contributions. With increasing delay time, the switching-current spectrum gradually evolves into three distinguishable peaks, which become clearly observable after 200 s, providing direct evidence for the coexistence of multiple domain populations.

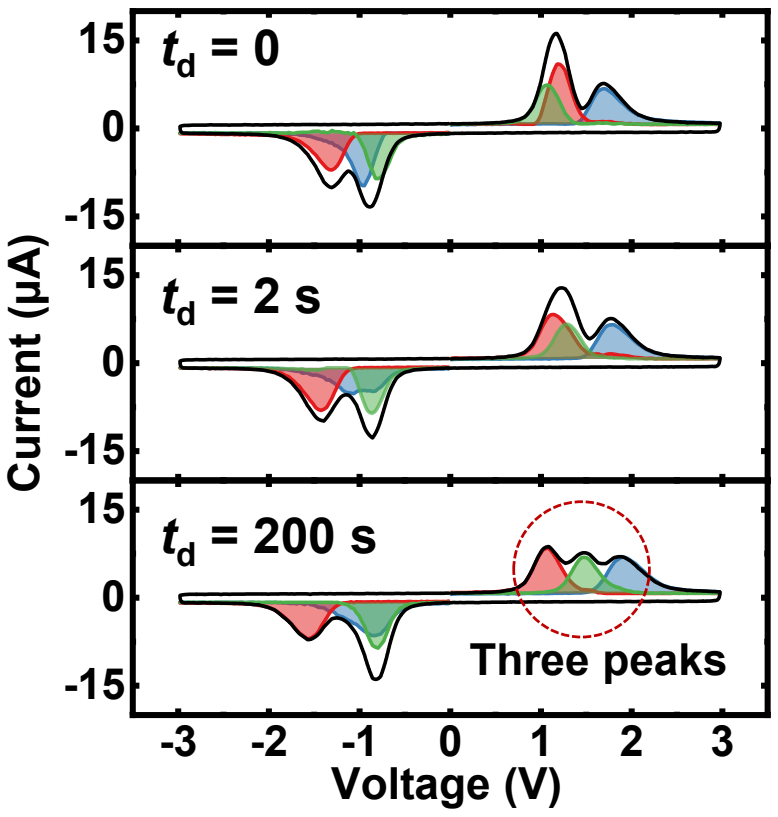

Fig. 6 *I-V* curves measured after electrical cycling with different delay times, with the contributions of distinct domain populations separated using the domain-decoupling method. The measurement after a 200 s delay clearly shows three individual current peaks.

## IV. Conclusion

In this work, we propose a three-domain model, consisting of switchable domains $\mathbf{P_d}$ and unswitchable domains $\mathbf{P_\uparrow}/\mathbf{P_\downarrow}$, to investigate heterogeneous domain populations in ferroelectric films. Based on this model, a novel domain decoupling method is developed to independently extract and quantify the responses of $\mathbf{P_\uparrow}$, $\mathbf{P_\downarrow}$, and $\mathbf{P_d}$. The proposed approach enables quantitative analysis of domain evolution and provides a useful tool for investigating reliability-related phenomena in ferroelectric devices.